# Using statistical smoothing to date medieval manuscripts[*]

### Andrey Feuerverger[1], Peter Hall[2], Gelila Tilahun[3] and Michael Gervers[4]

*University of Toronto, University of Melbourne, University of Toronto and University of Toronto*

**Abstract:** We discuss the use of multivariate kernel smoothing methods to date manuscripts dating from the 11th to the 15th centuries, in the English county of Essex. The dataset consists of some 3300 dated and 5000 undated manuscripts, and the former are used as a training sample for imputing dates for the latter. It is assumed that two manuscripts that are "close", in a sense that may be defined by a vector of measures of distance for documents, will have close dates. Using this approach, statistical ideas are used to assess "similarity", by smoothing among distance measures, and thus to estimate dates for the 5000 undated manuscripts by reference to the dated ones.

## 1. Introduction

The problem of searching for, and comparing, documents on the world wide web has motivated the development of techniques for measuring the "relationships" among documents. These methods include approaches based on formal measures of distance (see e.g. the collections of papers edited by Berry [1, 2] and Djeraba [6]), as well as more statistical techniques (see e.g. Cutting et al. [5]). A brief review of the literature, in a statistical context, has been given by Feuerverger et al. [7]. The present article builds on the latter paper, by giving relatively detailed accounts of the application of kernel-smoothing methods to the dating, or calendaring as it is often called, of medieval manuscripts.

These manuscripts are charters written between the 11th and the 15th centuries. They relate to property holdings or transfers in the county of Essex, England. Many were taken from entries in the *Hospitaller Cartulary* of 1442 and are essentially land deeds involving the Order of the Hospital of St John of Jerusalem (Gervers [11]). The Order, which works internationally today in healthcare, originated in the 11th century as a monastic brotherhood caring for pilgrims in the Holy Land. The dated manuscripts are part of a database assembled at the University of Toronto in

[*]Supported in part by the Natural Sciences and Engineering Research Council of Canada.
[1]Department of Statistics, University of Toronto, 100 St. George Street, Toronto, Ontario, Canada M5S 3G3, e-mail: andrey@utstat.toronto.edu
[2]Department of Mathematics and Statistics, University of Melbourne, Melbourne, VIC 3010, Australia, e-mail: p.hall@ms.unimelb.edu.au
[3]Department of Statistics, University of Toronto, 100 St. George Street, Toronto, Ontario, Canada M5S 3G3, e-mail: gelila@utstat.toronto.edu
[4]Department of History ETC., University of Toronto, 100 St. George Street, Toronto, Ontario, Canada M5S 3G3, e-mail: 102063.2152@compuserve.com

connection with the DEEDS (Documents of Essex England Data Set) project, and are discussed by Gervers [10, 12, 13].

In this paper we interpret the dates of undated documents as missing components of random data vectors of indeterminate length, and impute them using nonparametric, kernel-based regression in which the explanatory variables are interdocument distances. Sections 2, 3, 4 and 5 respectively discuss measures of distance, methodology for smoothing empirical distances, applications of these techniques to manuscript data, and theory relating to Section 3.

## 2. Shingles, resemblances and distances

Mathematical formalisation of a manuscript proceeds as follows. Remove all punctuation from the manuscript. The document that remains is a sequence of $n$, say, words, with repetitions counted as different words. Write this sequence as $\mathcal{M} = (w_1, \ldots, w_n)$. A consecutive sequence of $k$ words, i.e. $S = (w_{t+1}, w_{t+2}, \ldots, w_{t+k})$ where $0 \leq t \leq n-k$, is called a *shingle* of order $k$. Let $\mathcal{S}_k(\mathcal{M}) = \{S_{k1}, \ldots, S_{k,N(k)}\}$, where $1 \leq N(k) \leq n-k+1$, denote the set of distinct shingles of order $k$ obtainable from the manuscript $\mathcal{M}$.

If $\mathcal{M}_i$ and $\mathcal{M}_j$ are two manuscripts then the mathematical intersection of $\mathcal{S}_k(\mathcal{M}_i)$ and $\mathcal{S}_k(\mathcal{M}_j)$ is the set of different shingles of order $k$ that are contained in both manuscripts. Broder et al. [3] and Broder [4] introduced the notion of the $k$th order *resemblance distance*, $d_k(i,j)$, between $\mathcal{M}_i$ and $\mathcal{M}_j$. It is the proportion of shingles, out of the set of all $k$th order shingles in $\mathcal{M}_i$ and $\mathcal{M}_j$, that are not contained in both $\mathcal{M}_i$ and $\mathcal{M}_j$. It can be defined mathematically as

$$d_k(i,j) = 1 - \frac{\|\mathcal{S}_k(\mathcal{M}_i) \cap \mathcal{S}_k(\mathcal{M}_j)\|}{\|\mathcal{S}_k(\mathcal{M}_i) \cup \mathcal{S}_k(\mathcal{M}_j)\|},$$

where $\|\mathcal{S}\|$ denotes the number of elements of a finite set $\mathcal{S}$. We shall denote by $\mathrm{res}_k(i,j) = 1 - d_k(i,j)$ the $k$th order resemblance.

## 3. Smoothing step

We work with $r$-vectors of resemblance distances. In particular, the $k$th component of the vector describing the distance between $\mathcal{M}_i$ and $\mathcal{M}_j$ is $d_k(i,j)$, for $1 \leq k \leq r$. Thus, it is assumed that resemblance distances, of orders up to the $r$th, capture the principal ways in which manuscripts differ. There may, however, be significant information from other, ordered variables such as document length or the simple frequencies of certain key words. These could also be incorporated into our smoothing algorithm, by making obvious changes to the methodology discussed below. Categorical variables such as document "type", for example whether the document is a marriage contract or a land deed (in the context of Latin manuscripts discussed in Section 4), are arguably best included by smoothing within the respective type.

If manuscript $\mathcal{M}_i$ is undated, and manuscripts $\mathcal{M}_j$, for $j \in \mathcal{J}$, have respective known dates $t_j$, then the kernel weight applied to $\mathcal{M}_j$, based on its nearness to $\mathcal{M}_i$, is

$$a(i,j) = a(i,j \mid h_1, \ldots, h_r) = \prod_{k=1}^{r} K\{d_k(i,j)/h_k\},$$

where $K$ denotes a nonnegative, nonincreasing function defined on the positive half-line, and $h_1, \ldots, h_r$ are bandwidths. Our estimator, $\hat{t}_i$ of the date $t_i$ of $\mathcal{M}_i$ is

$$\hat{t}_i = \arg\min \left\{ \sum_{j \in \mathcal{J}} (t_j - t)^2 \, a(i,j) \right\} = \left\{ \sum_{j \in \mathcal{J}} t_j \, a(i,j) \right\} \Big/ \left\{ \sum_{j \in \mathcal{J}} a(i,j) \right\}. \quad (3.1)$$

Theoretical properties of this estimator will be discussed in Section 5.

There is precedent in the field of manuscript dating for using robust methods. In this context one could minimise the loss function $\sum_{j \in \mathcal{J}} \Psi(t_j - t) \, a(i,j)$, where $\Psi$ is a positive, symmetric, "cup shaped" function with its unique minimum at the origin and increasing no faster than linearly in its tails. This leads to the estimator $t = \hat{t}_i$ that solves

$$\sum_{j \in \mathcal{J}} \psi(t_j - t) \, a(i,j) = 0 \,,$$

where $\psi = \Psi'$. Taking $\Psi(u) \equiv |u|$ we obtain the local median. See Härdle and Gasser [14] for discussion of robust nonparametric methods in a univariate setting.

To choose bandwidths, define

$$\hat{t}_{j'} = \hat{t}_{j'}(h_1, \ldots, h_r) \equiv \arg\min_{t} \sum_{j \in \mathcal{J}, \, j \neq j'} (t_j - t)^2 \, a(j', j \,|\, h_1, \ldots, h_r) \,,$$

$$(\hat{h}_1, \ldots, \hat{h}_r) = \arg\min_{(h_1, \ldots, h_r)} \sum_{j' \in \mathcal{K}} \left\{ t_{j'} - \hat{t}_{j'}(h_1, \ldots, h_r) \right\}^2 \,,$$

where $\mathcal{K}$ is the union, over $1 \leq k \leq r$, of the set of all $j \in \mathcal{J}$ such that $d_k(i,j)$ is among the $m$ largest values of that quantity. Here $m$ would give an appropriately small fraction of the total number of dated manuscripts. Our empirical choice of the bandwidth vector is $(\hat{h}_1, \ldots, \hat{h}_r)$. This method is a form of predictive cross-validation. See Feuerverger et al. [7] for discussion of refinements.

## 4. Application to Latin manuscript data

### *4.1. The dataset, and approaches to calendaring*

At the time of writing this paper, the set of dated manuscripts consisted of 3353 *charters* written in the 11th, 12th, 13th, 14th and 15th centuries. In addition there were some 5000 undated manuscripts. An example of a dated manuscript is given in the Appendix.

The term "dated" above is used in an informal sense, and does not imply that a dated manuscript always had an exact date written upon it. Indeed, the majority of dated manuscripts are calendared by internal evidence, and inaccuracies are sometimes present. Gervers [12] details the nature and potential size of these errors; discrepancies of the order of several generations are possible. Witness names are one source of internal evidence, but identical witness names appear on manuscripts that could not possibly have been witnessed by the same person, and sometimes on manuscripts dated 100 years or more apart. See, for example, Rees ([15], p. xvii). Furthermore, witness names can be truncated, making them hard to identify reliably; or the names may be omitted altogether. Handwriting evidence can also be used for dating, but it too has drawbacks (e.g. Stenton [17], p. xxxii).

In cases where reliable internal dating is not possible, use can be made of the fact that the language, form and content of medieval manuscripts is constantly changing with time (e.g. Stenton [17]). This "dating by formulae" approach, as it is sometimes called, is outlined by Gervers [12, 13] in connection with the DEEDS manuscripts project. See also Fiallos [8, 9], who describes some algorithms for dating. Both Fiallos and Gervers refer to a "shingle" as a "word pattern" or "string"; the term "shingle" was introduced by Broder [4]. The techniques discussed by Gervers are substantially more interactive, and more demanding of expert historical knowledge, than automated-statistical approaches such as those suggested in the present paper. While this may enhance accuracy in some cases, it makes the methodology difficult to transfer to other applications.

### 4.2. Data analysis

Figure 1 shows a histogram giving the distributions of dates for the full dataset, of size 3353. The dates range from 1089 to 1466, and are seen to be more concentrated towards central values. By means of random sampling these documents were divided into three disjoint groups. The first group, consisting of 3034 documents, served as a training set; the second, of 167 documents, was used for validation; and the third, of 152 documents, was set aside to serve as a test set. The decision to set aside a validation subset, rather than use leave-one-out validation, was made to reduce computational labour. Below we report results obtained using resemblance distance; see Section 2 for a definition. Throughout we employed the exponential kernel, $K(x) = e^{-x}$.

We experimented with shingles of sizes one, two and three, as well as with all combinations of these sizes. With every such combination we varied the value of $m$, introduced in Section 3 and defined as the number of "closest" charters to be included in the weighted estimation procedure, over the range $m = 5, 10, 20$ and 50. Optimisation over $h$ was done by searching over a fine grid. The "shingling" of charters into lists of unique sequences of consecutive words, the matching of shingles among charters, the computation of distances between them, and the estimation of dates, were carried out employing C and standard UNIX commands.

Using shingle sizes greater than three was ruled out, to limit the amount of computation required and also in light of the findings reported below. In this section, so as to describe the effects of different shingle sizes, we shall present results in the setting of univariate smoothing over resemblance distances. That is, we took $r = 1$ in the setting of Section 3, although that single component was any one of $d_1(i,j)$, $d_2(i,j)$ and $d_3(i,j)$; the particular value will be made clear in discussion. Taking $r = 3$ produces results which have properties resembling an "average" of those discussed below.

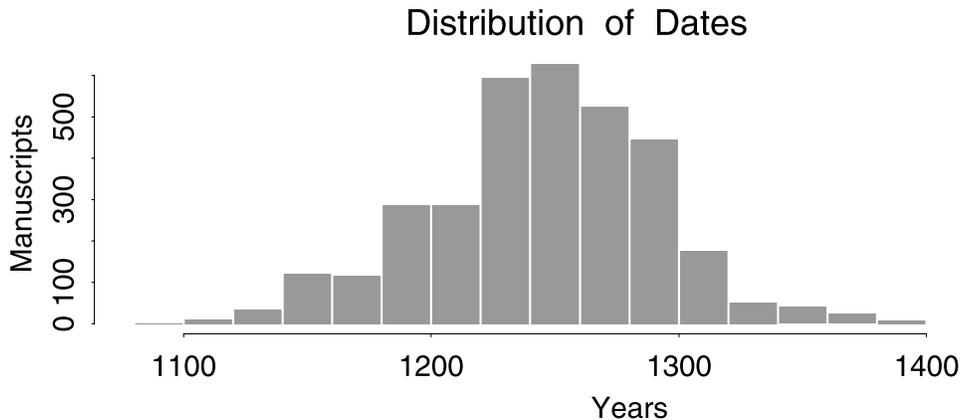

FIG 1. *Histogram of the dates for the complete dataset.*

To convey a broad impression of statistical properties of the data we mention that the total number of resemblance values among the validation and training manuscripts was $167 \times 3034 = 506{,}678$. The mean values were 0.083, 0.014 and 0.0042 for shingles of size $k = 1$, 2 and 3, respectively. The pairwise correlations between the 506,678 resemblance values, for resemblances based on shingle sizes $k = 1$, 2 and 3, were 0.93 between shingle sizes 2 and 3, 0.88 between shingle sizes 1 and 2, and 0.76 between shingle sizes 1 and 3. Resemblance values can occasionally be quite large; they exceeded 0.5 a total of 14, 7 and 4 times among the resemblances based on shingle sizes 1, 2 and 3, respectively. Note too that the mean year of the training documents was 1245.8. If this value were used as the date estimate for each document in the validation set, the mean absolute error would be 36.6 years.

For shingles of size $k = 1$, and using resemblance distance, the optimal value of $(m, h)$ was found to be approximately $(10, 9.0 \times 10^{-3})$. (The small values for bandwidth, here and below, reflect small values of resemblance.) These choices resulted in an average absolute difference between true and estimated dates for charters, within the validation set, of 13.1 years. When only the closest $m = 5$ charters were used the average error only increased slightly, to 13.2 years, while $m = 20$ and $m = 50$ resulted in average absolute errors of 13.2 and 13.3 years, respectively, when each was used in conjunction with its corresponding optimal bandwidth.

As can be seen, the results are quite robust against choice of $m$. Note, however, that for a particular value of $m$, and for a given charter in the validation set, it can occasionally happen that there are fewer than $m$ charters (in the training set) that have nonzero resemblance to it. When that occurred, only those charters having nonzero resemblances were included in the estimation procedure. In this sense the effective value of $m$ could occasionally be smaller than its nominal value, particularly when $m$ was large.

The results are also robust against varying $h$. For example, in the optimal case, $m = 10$, the average error stayed below 13.2 years provided $h$ remained in the range $(6.9 \times 10^{-3}, 1.1 \times 10^{-2})$. In these instances, as in other results cited below, the mean absolute error functions were invariably well behaved as $h$ and $m$ varied, and had clearly defined (if somewhat broad) minima. No instances of separated multiple minima were discovered during our experimentation.

Figure 2 shows a grey-scale "image" plot of shingle-size $k = 2$ resemblances between the 167 validation manuscripts (on the vertical axis) and the training manuscripts (on the horizontal axis). In producing this plot, the manuscripts on each axis were first ordered from earliest date to latest, and resemblance values exceeding 0.3 were set equal to 0.3, while values below 0.1 were set equal to zero. The training manuscripts were then grouped, with five consecutive manuscripts in each of 606 groups, and the resemblance values were averaged for each validation manuscript within each training group. These values were then normalized so that the maximum value for each validation manuscript was equal to one. Finally, resulting values at or below 0.8 were set equal to "white," while the value 1.0 was set equal to "black;" a linear grey scale was used between these values. The roughly diagonal character of this display mirrors the tendency for documents to have higher resemblance values with other documents of comparable dates. The wide scatter

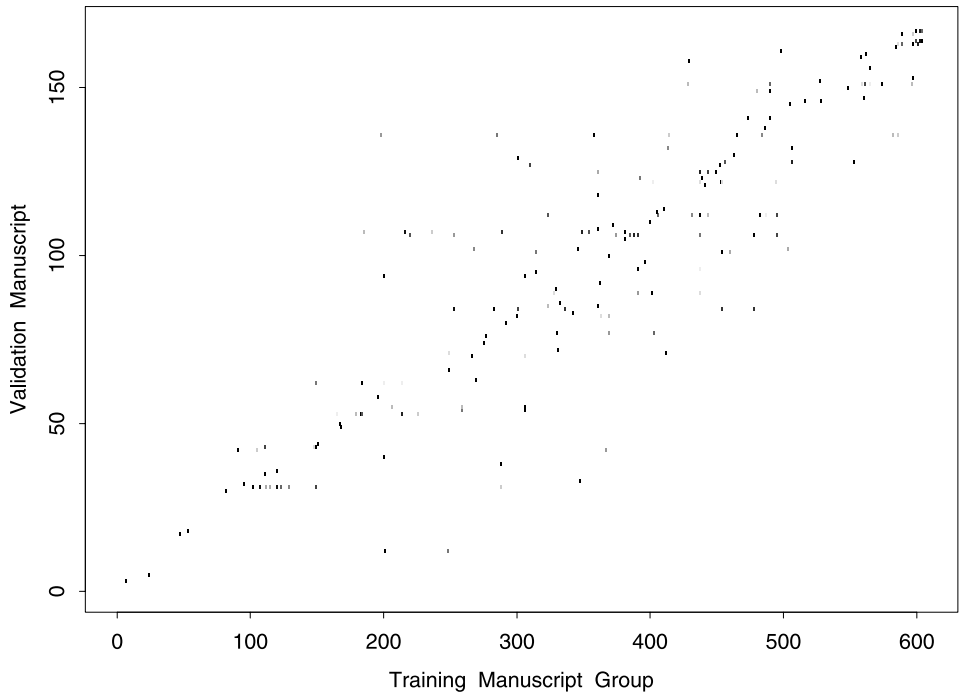

FIG 2. *Grey-scale plot of shingle-size 2 resemblances between the validation and training manuscripts.*

evident in the display also serves to emphasize the inherent difficulty of the problem.

For shingles of size $k = 2$, and using the resemblance measure, the optimal value of $(m, h)$ was $(5, 6.7 \times 10^{-3})$, giving a mean absolute error of 11.1 years. The mean absolute error stayed below 11.2 years provided $h$ remained in the range $(5.1 \times 10^{-3}, 9.5 \times 10^{-3})$. When $m$ increased to 10 or to 50 the best mean absolute error increased to 11.6 or 11.7 years, respectively. Once again, results are seen to be robust against choice of $m$.

Analogously, for shingles of size $k = 3$ the optimal value of $(m, h)$ was $(10, 2.0 \times 10^{-3})$, giving a mean absolute error of 12.1 years. For $m = 5$ and $m = 50$ the error was 12.4 and 12.2 years, respectively.

We also experimented with using resemblance measures for pairs of shingle sizes, employing bivariate kernels that were products of univariate kernels. We found that whenever shingle size $k = 2$ was included in a pair the optimisation attempted to eliminate the effect of the other shingle size (1 or 3) by assigning to it a relatively high (or even infinite) bandwidth. In consequence the results were virtually identical to those achieved using shingle size 2 alone: the minimum error achieved using shingle sizes 1 and 2 simultaneously was 11.2 years, and was achieved with $m = 5$, while for shingle sizes 2 and 3 together it was 11.1 years and also occurred with $m = 5$. By way of contrast, for shingle sizes 1 and 3 together the optimal error was 11.8 years and was attained with $(m, h_1, h_3) = (10, 0.0024, 0.05)$, where $h_j$ denotes the bandwidth applied to shingle size $j$. (Note that when more than one shingle size is used, the effective value of $m$ typically is somewhat increased, since the union is taken of the $m$ closest documents for each shingle size.)

The result for using all three shingle sizes (1, 2 and 3) simultaneously was simi-lar to that when using shingle size 2 alone, or using pairs of shingle sizes including size 2: minimum mean absolute error was achieved with $m = 5$, and was 11.2 years. Considering this and the previous results, it is seen that the best among the procedures discussed so far is to use only the resemblance distance based on shingles of size 2, taking $(m, h) = (5, 6.7 \times 10^{-3})$. Finally, to obtain independent verification of the claimed error rate, this procedure was applied to the 152 documents in the test set; the mean absolute error was found to be 12.2 years. The apparent difference between the error rates for the validation and test subsets is consistent with the bias due to having selected the best of several procedures, with the somewhat small sample sizes of these subsets, and with the fact that (due to random sampling) the manuscripts in these subsets had slightly different distributions of dates and wordcounts. Figure 3 shows the true dates for manuscripts in the test set, together with their estimated dates. (The zero error line is also superimposed.) Very slight bias effects are discernible at the edges due to the fact that for such manuscripts closest matches cannot exist at more extreme dates. It is also seen that manuscripts near the central date ranges are estimated slightly more accurately, in part owing to the availability of many more training manuscripts in the central date ranges (though there is also some countervailing influence, since the presence of more documents at a certain date slightly biases the selection of closest fitting manuscripts to that date). The size of a document being dated was also found to have a modest influence on how accurately it could be dated. Very large documents appear to have been dated somewhat more accurately than others, but very small documents were not, in general, dated inaccurately; indeed, the claimed mean absolute error rate appears to be quite generally applicable overall. (Versions of Figure 3 in which dot

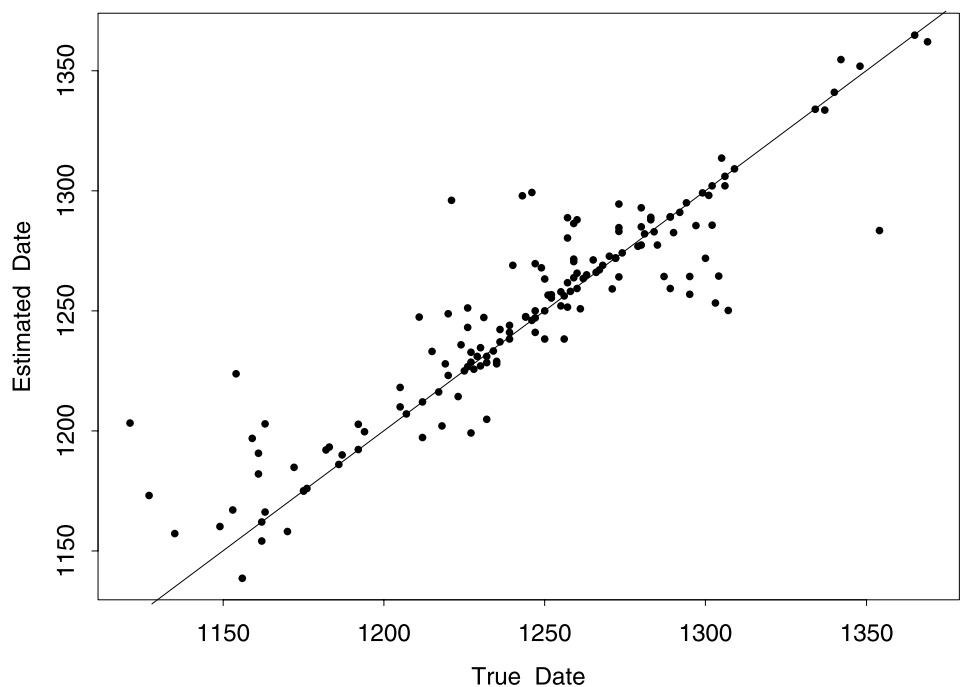

FIG 3. *True and estimated dates for manuscripts in the test set with the zero error line superimposed.*

sizes varied to reflect manuscript size did not prove to be informative.)

Finally, we mention that similar detailed numerical studies were also conducted for other definitions of distance discussed by Feuerverger et al. [7]. In each of these cases, the results obtained were broadly similar to those for resemblance distance, with shingles of size 2 alone again being the apparently optimal choice on which to base the estimation procedures. For these distances, the optimal mean absolute errors obtained all turned out to be slightly larger than that achieved using the resemblance distance.

## 5. Theoretical properties of kernel imputation

Let $\mathcal{M}_0$ be a particular undated manuscript, written at a time $t_0$, and let $(\mathcal{M}_0, \mathcal{M})$ denote the value of the pair $(\mathcal{M}_i, \mathcal{M}_j)$ when $\mathcal{M}_i = \mathcal{M}_0$ and $\mathcal{M}_j = \mathcal{M}$ is a randomly chosen, dated manuscript. Write $\Delta_\ell$ for the corresponding value of $d_\ell(i,j)$. Although the numbers of words in both a manuscript and a dictionary are finite, they are potentially so large that values of $d_\ell(i,j)$ are virtually distributed in the continuum. Therefore it is appropriate to model the joint distribution of the date $T$ of a manuscript $\mathcal{M}$, and the distances $\Delta_\ell$, by that of a vector $(T, \Delta_1, \ldots, \Delta_r)$ distributed in the continuum within the region $(0, \infty) \times [0,1]^r$.

In this setting our kernel method for estimating the unknown date $t_0$ is consistent if the following two assumptions are satisfied: (a) the mean value of $T$, evaluated when the distances $\Delta_1, \ldots, \Delta_r$ are in arbitrarily small neighbourhoods of 0, converges to $t_0$ as those neighbourhoods shrink to 0; and (b) the mean square of $T$, evaluated in the same context, remains bounded. The first of these conditions is one of "asymptotic unbiasedness" of the date of a random manuscript $\mathcal{M}$, as the distances between $\mathcal{M}$ and the fixed manuscript $\mathcal{M}_0$ converge to 0. The second condition is a common assumption of finite variance.

Respectively, these two conditions may be stated formally as

$$\frac{E\{T\,I(\Delta_1 \leq \delta_1, \ldots, \Delta_r \leq \delta_r)\}}{P(\Delta_1 \leq \delta_1, \ldots, \Delta_r \leq \delta_r)} \to t_0 \tag{5.1}$$

as $\delta_1, \ldots, \delta_r \to 0$, and

$$\frac{E\{T^2\,I(\Delta_1 \leq \delta_1, \ldots, \Delta_r \leq \delta_r)\}}{P(\Delta_1 \leq \delta_1, \ldots, \Delta_r \leq \delta_r)} \tag{5.2}$$

remains bounded as $\delta_1, \ldots, \delta_r \to 0$. Assume in addition that for each $c > 1$,

$$\limsup_{\delta_1, \ldots, \delta_r \to 0} \frac{P(\Delta_1 \leq c\,\delta_1, \ldots, c\,\Delta_r \leq \delta_r)}{P(\Delta_1 \leq \delta_1, \ldots, \Delta_r \leq \delta_r)} < \infty\,;$$

that the kernel $K$ is bounded, continuous, compactly supported and nonincreasing on the positive real line; that $K(x_0) > 0$ for some $x_0 \geq 0$; that the dated manuscripts $\{\mathcal{M}_j, j \in \mathcal{J}\}$ are independent and identically distributed as $\mathcal{M}$; that the number $N(\mathcal{J})$ of elements of $\mathcal{J}$ is allowed to increase to infinity; and that at the same time as $N(\mathcal{J})$ increases, the bandwidths $h_1, \ldots, h_r$ decrease to 0, but so slowly that

$$N(\mathcal{J})\,P(\Delta_1 \leq h_1, \ldots, \Delta_r \leq h_r) \to \infty\,. \tag{5.3}$$

Let C denote the set of all conditions stated in this paragraph.

**Theorem 5.1.** *If* C *holds then the estimator $\hat{t}_i$ defined at* (2.1) *converges in probability to the true date of the undated manuscript $\mathcal{M}_0$.*

Under more refined conditions, properties of bias and variance may be derived. In particular it may be shown that variance is generally of order $\{N(\mathcal{J})\,h_1\ldots h_r\}^{-1}$ and bias of order $\max(h_1^2,\ldots,h_r^2)$, and that the estimator is asymptotically Normally distributed.

To derive the theorem, note that for each $\epsilon > 0$, any kernel $K$ that satisfies conditions C may be sandwiched between two kernels $K_1$ and $K_2$, with the properties: (a) $K_1 \leq K \leq K_2$, (b) $K_2(x) - K_1(x) \leq \epsilon$ for all $x$, and (c) $K_1$ and $K_2$ are each expressible as finite, positive linear combinations of functions of the form $L(x) = I(0 < x \leq c)$, where $c > 0$. We first derive the theorem in the case $K = L$.

Define $A = \sum_{j \in \mathcal{J}} t_j\, a(i,j)$, $B = \sum_{j \in \mathcal{J}} a(i,j)$, $\delta_\ell = c\, h_\ell$, $\delta = (\delta_1,\ldots,\delta_r)$ and $\pi(\delta) = P(\Delta_1 \leq \delta_1, \ldots, \Delta_r \leq \delta_r)$. Then by (5.1) and (5.2) we have in the case $K = L$,

$$E(A) = N(\mathcal{J})\, E\{T\, I(\Delta_1 \leq \delta_1, \ldots, \Delta_r \leq \delta_r)\} = \{t_0 + o(1)\}\, N(\mathcal{J})\, \pi(\delta)\,,$$

$$\mathrm{var}(A) \leq N(\mathcal{J})\, E\{T^2\, I(\Delta_1 \leq \delta_1, \ldots, \Delta_r \leq \delta_r)\} = O\{N(\mathcal{J})\, \pi(\delta)\}\,,$$

$E(B) = N(\mathcal{J})\, \pi(\delta)$ and $\mathrm{var}(B) \leq N(\mathcal{J})\, \pi(\delta)$. It follows from these results and (5.3) that $A/\{N(\mathcal{J})\, \pi(\delta)\} \to t_0$ and $B/\{N(\mathcal{J})\, \pi(\delta)\} \to 1$, both convergences being in probability. Therefore $A/B \to t_0$ in probability, as had to be shown.

Finally we treat a more general kernel satisfying conditions C. Using the properties noted two paragraphs above we may deduce from the results for $K = L$ in the previous paragraph that for each $\epsilon > 0$,

$$E(B)\,\{t_0 - \epsilon + o(1)\} \leq E(A) \leq E(B)\,\{t_0 + \epsilon + o(1)\}\,,$$

$$\mathrm{var}(A) + \mathrm{var}(B) = O\{E(B)\}\,,$$

$$E(B) \asymp N(\mathcal{J})\, P(\Delta_1 \leq h_1, \ldots, \Delta_r \leq h_r)\,.$$

Again these results imply the desired property $A/B \to t_0$.

## Appendix A: Typical dated manuscript from database

Document 00640214, as it appears in the database, is given below. The manuscript's date, 1237 AD, is part of its header. All punctuation marks have been removed, and numbers (in Roman numerals) are given between exclamation marks. Each number is replaced by simply "#" before shingling, so different numbers are not distinguished. However, shingling distinguishes capitalised from non-capitalised words; for example, "regis" is regarded as different from "Regis".

00640214    1237

*Haec est finalis concordia facta in curia domini regis apud Westmonasterium a die S Johannis Baptistae in !xv! dies anno regni regis Henrici filii regis Johannis !xxi! coram Roberto de Lexinton Willelmo de Eboraco Ada filio Willelmi Willelmo de Culewurth justitiariis et aliis domini regis fidelibus tunc ibi praesentibus inter Johannem Baioc quaerentem et Robertum Sarum episcopum et capitulum deforciantes per Radulfum de Haghe positum loco ipsorum ad lucrandum vel perdendum de advocatione ecclesiae de Waye Bayouse unde assisa ultimae praesentationis summonita fuit inter eos in eadem curia scilicet quod praedictus T recognovit advocationem praedictae ecclesiae cum pertinentiis esse jus ipsorum episcopi et capituli et ecclesiae suae Sarum ut illam quam idem episcopus et capitulum Sarum habent de dono Alani de Baiocis patris praedicti Johannis cujus haeres ipse est et idem episcopus et capitulum praedictum concesserunt pro se ob successoribus suis eidem Johanni ut eidem ecclesiae quotiescunque tota vita ipsius eam vacare contigerit possit idoneam*

*personam praesentare ita quod quicunque pro tempore fuerit persona ejusdem ecclesiae ad praesentationem ipsius Johannis reddet singulis annis praedictis episcopo et capitulo sex marcas argenti de praedicta ecclesia apud Sarum nomine pensionis scilicet ad festum S Michaelis !xx! solidos ad Natale Domini !xx! solidos ad Pascha !xx! solidos ad nativitatem beati Johannis Baptistae !xx! solidos et post decessum ipsius Johannis advocatio praedictae ecclesiae cum pertinentiis remanebit praedictis episcopo et capitulo Sarum et eorum successoribus quieta de haeredibus ipsius Johannis in perpetuum Et praeterea idem episcopus et capitulum praedictum concesserunt pro se et successoribus suis quod ipsi de caetero invenient unum capellanum divina celebrantem singulis diebus anni in capella beati Johannis sita infra parochiam de Waye pro anima praedicti Johannis et pro animabus haeredum suorum et antecessorum suorum et pro cunctis fidelibus in perpetuum et idem episcopus et capitulum praedictum et successores sui invenient ornamenta libros et luminaria sufficientia in eadem capella in perpetuum*

# References

[1] BERRY, M. W. (2001). *Computational Information Retrieval*. SIAM, Philadelphia. MR1861811

[2] BERRY, M. W. (2003). *Survey of Text Mining: Clustering, Classification, and Retrieval.* Springer, New York.

[3] BRODER, A. Z., GLASSMAN, S. C., MANASSE, M. S. AND ZWEIG, G. (1997). Syntactic clustering of the web. SRC Technical Note No. 1997-015, Digital Equipment Corporation. In *Proceedings of the Sixth International World Wide Web Conference* 391–404.

[4] BRODER, A. Z. (1998). On the resemblance and containment of documents. In *1997 International Conference on Compression and Complexity of Sequences (SEQUENCES '97)*, June 11–13 1997, Positano, Italy, 21–29. IEEE Computer Society, Los Alamitos, California.

[5] CUTTING, D. R., KARGER, D. R., PEDERSEN, J. O. AND TUKEY, J. W. (1992). Scatter/gather: a cluster-based approach to browsing large document collections. In *Proc. Fifteenth Annual International ACM SIGIR Conference on Research and Development in Information Retrieval*, Copenhagen, Denmark, June 21–24 1992 (N. J. Belkin, P. Ingwersen and A. M. Pejtersen, eds.) 318–329. Association for Computing Machinery, New York.

[6] DJERABA, C. (2002). *Multimedia Mining – A Highway to Intelligent Multimedia Documents*. Kluwer, Boston.

[7] FEUERVERGER, A., HALL, P., TILAHUN, G. AND GERVERS, M. (2005). Distance measures and smoothing methodology for imputing features of documents. *J. Statist. Graph. Statist.* **14** 255–262. MR2160812

[8] FIALLOS, R. (2000a). An overview of the process of dating undated medieval charters: latest results and future developments. In *Dating Undated Medieval Charters* (M. Gervers, ed.) 37–48. Boydell Press, Woodbridge, UK.

[9] FIALLOS, R. (2000b). Procedure for dating undated documents using a rational database. Manuscript.

[10] GERVERS, M. (1989). The textile industry in Essex in the late 12th and 13th centuries: A study based on occupational names in charter sources. *Essex Archaelogy and History* **20** 34–73.

[11] GERVERS, M. (1982, 1996). *The Cartulary of the Knights of St. John of Jerusalem in England*, Parts 1, 2. Oxford Univ. Press, London.

[12] GERVERS, M. (2000a). The DEEDS project and the development of a computerised methodology for dating undated English private charters of the twelfth

and thirteenth centuries. In *Dating Undated Medieval Charters* (M. Gervers, ed.) 13–35. Boydell Press, Woodbridge, UK.
[13] GERVERS, M. (2000b). The dating of medieval English private charters of the twelfth and thirteenth centuries. Manuscript.
[14] HÄRDLE, W. and GASSER, T. (1984). Robust nonparametric function fitting. *J. Roy. Statist. Soc.* Ser. B **46**, 42–51. MR0745214
[15] REES, U. (1975). *The Cartulary of Shrewsbury Abbey* **1**. Aberystwyth.
[16] RABIN, M. O. (1981). Fingerprinting by random polynomials. Report TR-15-81, Center for Research in Computing Technology, Harvard Univ.
[17] STENTON, F. M. (1922). *Transcripts of Charters relating to the Gilbertine Houses of Sixle, Ormsby, Catley, Bullington, and Alvingham. Publications of the Lincoln Record Society for 1920* **18**. Horncastle, UK.